\documentclass[superscriptaddress,amsmath,amssymb, reprint]{revtex4-1}

\usepackage{graphicx}
\usepackage{dcolumn}
\usepackage{bm}

\usepackage[utf8]{inputenc}
\usepackage[T1]{fontenc}
\usepackage{amsmath, amssymb}
\usepackage{newtxtext,newtxmath}
\usepackage{etoolbox}
\usepackage{upgreek}
\usepackage[version=4]{mhchem}
\usepackage{fix-cm}
\usepackage{booktabs}
\usepackage{hyperref}

\newcommand{\ISMO}{Institut des Sciences moleculaires d'Orsay, CNRS UMR 8214, Universite Paris-Saclay, 520 Rue Andre Riviere, 91400 Orsay}
\newcommand{\SOLEIL}{Synchrotron SOLEIL, L'Orme des Merisiers, St. Aubin BP 48, 91192 Gif-sur-Yvette, France}

\makeatother

\begin{document}

\title[Angle-resolved photoelectron spectroscopy of the DABCO molecule]{Angle-resolved photoelectron spectroscopy of the DABCO molecule probed with VUV radiation}

\author{Audrey Scognamiglio}
\affiliation{\ISMO}

\author{Lou Barreau}
\affiliation{\ISMO}

\author{Constant Schouder}
\affiliation{\ISMO}

\author{Denis Cubaynes}
\affiliation{\ISMO}

\author{Bérenger Gans}
\affiliation{\ISMO}

\author{Éric Gloaguen}
\affiliation{\ISMO}

\author{Gustavo A. Garcias}
\affiliation{\SOLEIL}

\author{Laurent Nahon}
\affiliation{\SOLEIL}

\author{Lionel Poisson}
\affiliation{\ISMO}
\email{lionel.poisson@universite-paris-saclay.fr}

\date{\today}

\begin{abstract}
We report a study of the diazabicyclo[2.2.2]octane (DABCO) molecule photoionized using VUV synchrotron radiation in combination with an ion–electron coincidence spectrometer. We determine accurately the adiabatic ionization energy to $7.199\pm0.006$~eV. Two vibrational progressions of DABCO cation ground state are resolved at $847~\text{cm}^{-1}\pm27~\text{cm}^{-1}$ and $1257~\text{cm}^{-1}\pm67~\text{cm}^{-1}$, which we assign to modes of $e'$ symmetry. Analysis of the photoelectron angular distribution shows that the anisotropy parameter depends on the vibrational excitation. This dependence of the $\beta$ parameter with the vibrational excitation is attributed to the scattering of the outgoing wavefunction mediated by high-lying Rydberg states.
\end{abstract}

\maketitle

\section{Introduction}
\begin{figure}[htbp]
    \centering
    \includegraphics[width=0.25\linewidth]{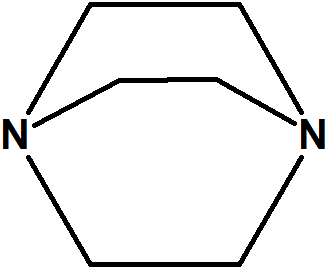}
    \caption{Structure of the DABCO molecule}
    \label{fig:DABCO}
\end{figure}

The diazabicyclo[2.2.2]octane (DABCO) molecule - equally named triethylenediamine - is a bicyclic organic compound with the molecular formula $\mathrm{N_2}$($\mathrm{C_2H_4}$)$_3$ (cf. Figure~\ref{fig:DABCO}), well-known in chemistry as a low-cost, basic organocatalyst, suitable for green-chemistry \cite{chakraborty_versatility_2023}. The DABCO molecule can be converted to a stable ammonium derivatives by alkylation of its nitrogen atoms \cite{maras_ring-opening_2012}. These DABCO-derived quaternary ammoniums are widely used in the design of ionic liquids \cite{turgula_third-generation_2020, soltanabadi_dft_2025, al_ans_ionic_2024}, a class of materials recognized for their broad applicability in areas including for example separation processes, electrochemical devices, and analytical techniques. 
The lack of $\mathrm{\pi}$ bonds implies the existence of energetically high excited states: typically the two lowest electronic excitations of the DABCO molecule are already of Rydberg character \cite{avouris_low-energy_1981, consalvo_dabco_1992}. Studies have focused on measuring and assigning the lower excited states of the DABCO molecule using multiphoton ionization \cite{parker_multiphoton_1978}, multiphoton ionization combined with fluorescence spectroscopy \cite{parker_multiphoton_1979} and ab-initio quantum calculations \cite{avouris_low-energy_1981}. The lower-lying excited state $\rm S_1$ corresponds to the excitation from the HOMO to the $\mathrm{\text{3s}}$ Rydberg orbital. Given that it has the same symmetry as the ground state ($\mathrm{A_1}'$), it is either accessible in a two-photon excitation or in a one-photon excitation via vibronic coupling \cite{consalvo_dabco_1992, consalvo_high-resolution_1993}. The second excited state $\rm S_2$ corresponds to the promotion of an electron from the HOMO to the $\mathrm{\text{3p}_{x,y}}$ Rydberg orbital, a state with symmetry $\mathrm{E}'$ accessible under one or two-photon excitation. Another state of symmetry $\mathrm{A_2}''$ accessible in a one-photon transition would correspond to the promotion of an electron to the Rydberg orbital $\mathrm{\text{3s}}$. This state that has a low transition dipole moment and is unobserved. It lies energetically nearby the $\rm S_2$ state \cite{smith_two-color_1984, pratt_photoionization_2002}. Higher lying Rydberg states have been measured using doubly resonant excitation and vibrational autoionization, also allowing the determination of the adiabatic ionization energy to 7.32~eV by fitting the Rydberg series \cite{boogaarts_high_1996}. Earlier multiphoton ionization studies allowed to resolve the Rydberg series from which the ionization threshold was calculated \cite{fujii_two-color_1983}.  Other publications report an adiabatic ionization energy of 7.197~eV \cite{smith_two-color_1984}, 7.23~eV \cite{parker_determination_1979}, and 7.2~eV \cite{parker_multiphoton_1981, parker_multiphoton_1979} measured using two-color laser photoionization spectroscopy. The adiabatic ionization threshold measured by photoelectron spectroscopy was found to be 7.195~eV \cite{pratt_photoionization_2002}. 
\newline
The vibrational structure of the $\rm S_0$, $\rm S_1$ and $\rm S_2$ states of DABCO in the gas phase has been thoroughly studied by means of fluorescence spectroscopy \cite{consalvo_dabco_1992,consalvo_high-resolution_1993,gonohe_one-photon_1982, parker_multiphoton_1979} or multiphoton ionization \cite{parker_multiphoton_1978,parker_multiphoton_1979}. IR and Raman spectra were recorded in solution \cite{ernstbrunner_free_1978, guzonas_raman_1988}, crystals or solid phase \cite{sauvajol_raman_1980, weiss_vibrational_1964, marzocchi_structure_1965} or gas phase \cite{marzocchi_structure_1965}. In a recent study performed by \citet{kovalenko_experimental_2012}, the IR and Raman vibrational spectra of the DABCO molecule in its ground state are compared to DFT based calculation. Most of the vibrational modes are associated either with \ce{C-N} stretching, $\mathrm{CH_2}$ bending or stretching, and to low-frequency cage torsion and deformation modes. Few publications focus on the fragmentation pattern observed following resonant population of the lower Rydberg vibronic states \cite{newton_two-color_1981, lichtin_laser_1980, parker_multiphoton_1981}. 
\newline 
Rydberg states usually have long lifetimes, however time-resolved photoelectron spectroscopy used to probe the dynamics of the $\rm S_2$ state has revealed the latter to be efficiently non-adiabatically coupled to the $\rm S_1$ state. The distortions of the $\rm S_1$ and $\rm S_2$ potentials are attributed to vibronic coupling and a Jahn-Teller distortion of the ground $\rm S_2$ state\cite{boguslavskiy_non-bornoppenheimer_2011}. Most of the earlier studies report a long fluorescence lifetime (about 1~$\upmu$s \cite{smith_two-color_1984,parker_multiphoton_1979}) of the DABCO molecule excited in the $\rm S_1$ state. The efficient population of two vibrational modes following intramolecular vibrational relaxation from the $\mathrm{S_1}$ state has been reported in a time-resolved femtosecond experiment \cite{poisson_unusual_2010}.
\newline
Literature on the neutral molecule is extensive, however only a few studies are dedicated to the cation. In this study, we employ a high-resolution, tunable VUV synchrotron source to perform valence photoelectron spectroscopy of the DABCO molecule. The first part of this article focuses on the cation vibrational structure observed by photoelectron spectroscopy. In the second part, we discuss unexpected features observed in the photoelectron angular distribution, namely a dependence of the anisotropy parameter on the vibrational excitation. This study is expected to enable future theoretical investigations of the anisotropy parameter of complex systems, starting from a case that is more straightforward to model.

% % EXPERIMENTAL METHODS 

\section{Experimental Methods}

The experiments were performed using the monochromatized, linearly polarized VUV light available at the DESIRS undulator-based beamline \cite{nahon_desirs_2012} of the French synchrotron facility SOLEIL. The high photon flux (typically close to $10^{13}$ photons per second between 7~eV and 12~eV), the tunability and high-resolution offered by the DESIRS beamline is particularly well suited for our studies. We used Krypton as a gas filter to remove the higher order harmonics of the undulator \cite{mercier_experimental_2000}. The monochromator calibration was realized by using the 
$\mathrm{Kr}\,5s\,$[3/2]$_{J=1}$ and the $\mathrm{Kr}\,5s\,$[1/2]$_{J=1}$ energy levels. The double imaging electron-ion coincidence spectrometer i2PEPICO, DELICIOUS III, which combines a velocity map imaging device (VMI) and a Wiley-Mc Laren time-of-flight spectrometer (TOF) was used for the detection of the photolectrons and photoions respectively \cite{garcia_delicious_2013}. 
The crystalline molecular sample (Sigma-Aldrich 98\%) was placed in the oven of the SAPHIRS molecular beam chamber and heated to temperatures between 40\textdegree~C and 90\textdegree~C. The resulting vapor was co-expanded with a rare gas or a mixture of rare gases through a $50~\upmu$m nozzle. We used either pure helium at 0.7~bar (data labeled $\rm A$), pure argon at 1.5~bar (data labeled $\rm B$), or a mixture of 20\% argon and 80\% helium at 2~bar (data labeled $\rm C$). The latter conditions were primarily set to facilitate cluster formation, which will be addressed in a forthcoming publication. In the present work, no cluster contribution was observed in either the photoelectron or the ion signal, particularly with respect to fragmentation. The supersonic beam passes through a 1~mm skimmer placed before the ionization chamber. The electrons and ions collected were analyzed in coincidence, so that the photolectron images are retrieved by filtering on the mass and on the translational energy. 
%This significantly limits the possible contribution of cluster dissociation in the monomer signal, given that cluster formation is possible along the supersonic expansion. 
The photoelectron spectra were obtained with the slow photoelectron spectrum (SPES) threshold method, which has shown to provide a good compromise between signal-to-noise ratio and resolution \cite{poully_photoionization_2010}. The photoelectron energy and angular distributions are retrieved using Abel inversion for a given species at each photon energy. This yields two 2D matrices—one for the energy distribution and one for the angular distribution—which are then integrated over the cationic states. In this work, we have used photolectrons up to 300~meV as a threshold for the integration. Because the electric field in the VMI region has the effect of lowering the ionization threshold, we have applied a correction factor \cite{barillot_influence_2017}:
\begin{equation}
    E_{\text{corr}}~[\text{cm}^{-1}] = -6.1212\,\sqrt{F~[\text{V/cm}]}
\end{equation}
The effective electric field $F$ inside the interaction region is given in \citet{garcia_delicious_2013} for the geometry of the DELICIOUS III detector:
\begin{equation}
    F~[\text{V/cm}] =  \frac{|V_{\text{rep}}|\,(1 - r) \, \alpha}{d}
\end{equation}
with $r=V_{\text{rep}}/V_{\text{ext}} = 0.71$, the usual VMI ratio \cite{eppink_velocity_1997}, $d=1.5~\text{cm}$ as in. \cite{eppink_velocity_1997}, and $\alpha=1.38603$ as in \cite{garcia_delicious_2013}  
This correction factor is of $6.21~\rm meV, 7.35~\rm meV$ and $10.76~\rm meV$ for measurements $\rm A$, $\rm B$ and $\rm C$, applied to all the values reported in this work. Additionally, the \textit{x}-axis of all spectra was corrected for visualization purposes. The features attributed to autoionizing Rydberg states, which are not affected by the Stark shift, are known with a precision of $\leq$~0.1 eV, which is 1 order of magnitude higher than the highest Stark shift calculated. We worked with an energy resolution of 0.04\% ($\approx~3~\text{meV}$, measurement $\rm A$), 0.21\% ($\approx~15~\text{meV}$, measurement $\rm B$) and 0.08\% ($\approx~6~\text{meV}$, measurement $\rm C$) of the photon energy\footnote{Values given between 7~eV and 8~eV}, with a step size of respectively 2.5~meV, 7~meV and 4~meV. Note that the resolution of the SPES will mostly be given by the electron resolution, while that of the total ion yield will only be given by the photon energy resolution. In the following, we describe our findings based on these three measurements, whose conditions were given above.

% %RESULTS AND DISCUSSION

\section{Results and Discussion}
\subsection{\label{sec:Vib-level}Vibrational spectrum}
\begin{figure}[ht]
    \centering
    \includegraphics[width=0.4\textwidth]{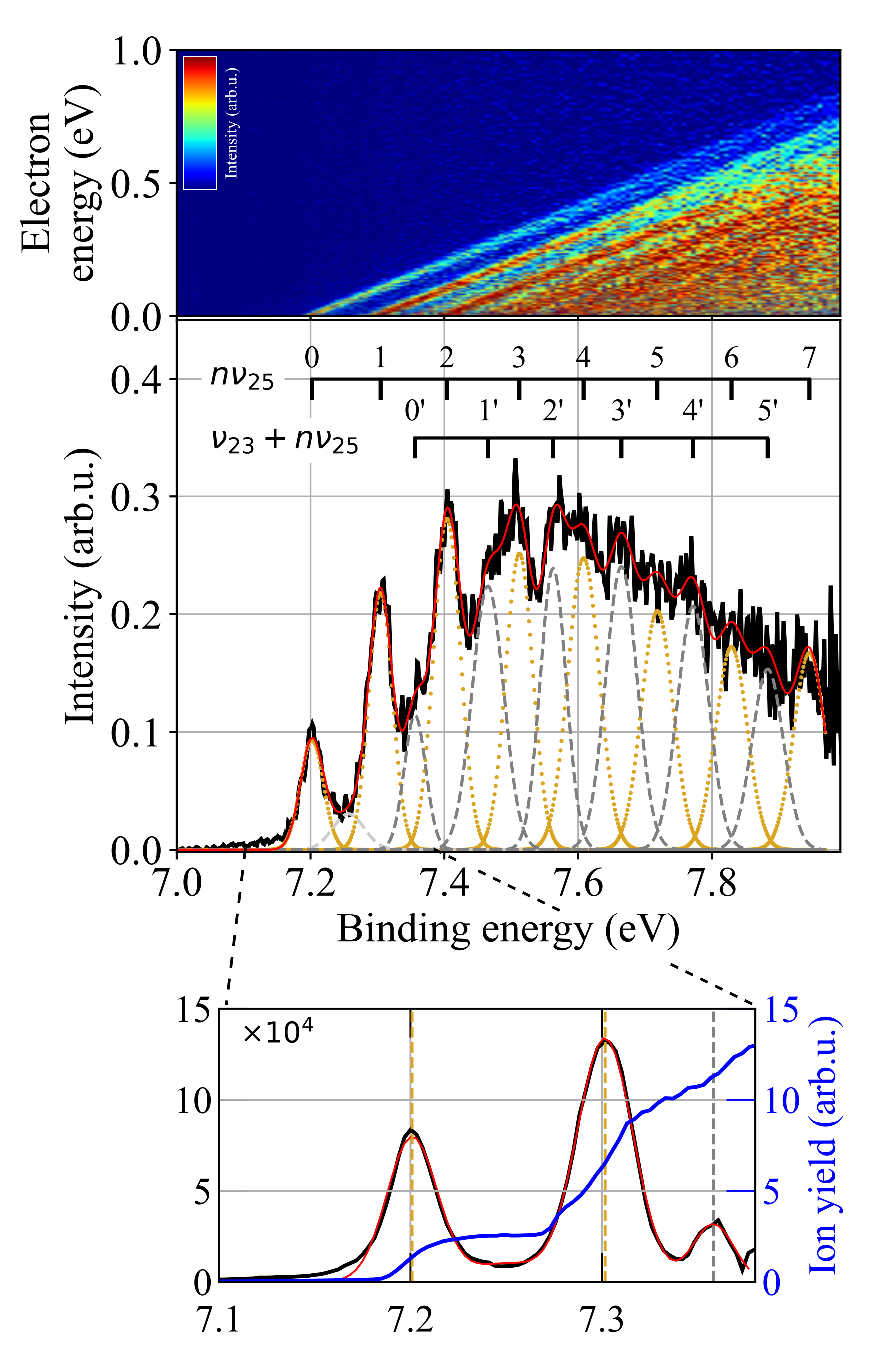}
    \caption{2D photoelectron matrix (upper panel) and SPES of DABCO (black trace, lower panel). The red solid line is the best fit to the data, composed of two Gaussian series (grey dashed and yellow dotted lines) assigned to two distinct vibrational progressions. Inset: photoelectron (black) and ion spectrum (blue) recorded near the ionization threshold. The grey and yellow dashed lines indicates the estimated position of the vibronic transitions. The \textit{x}-axis is corrected by the Stark shift of respectively $6.21~\rm meV$ and $10.76~\rm meV$ for the main figure and the inset.}
    \label{fig:SPES}
\end{figure}

The 2D photoionisation matrix and the SPES of the DABCO molecular ion are shown in Fig.~\ref{fig:SPES} upper and lower panels respectively. The 2D matrix (Fig.~\ref{fig:SPES}, upper panel) is built with the photoelectron spectrum obtained for each photon energy. Diagonal lines correspond to direct ionization and are the product of energy conservation. The SPES spectrum (Fig.~\ref{fig:SPES}, lower panel) is the integrated photoelectron signal over the first 300~meV of electron kinetic energy, which in that case, provided a good signal to resolution trade-off. The spectrum is fitted by a sum of Gaussian functions (red curve). The first colored band corresponds to the ionization threshold, and the successive bands are associated with the population of the vibrational states of the cation in its ground state. At high photon energies, many vibrational energy levels of different modes can be populated, thus resulting in spectral congestion, visible through the broadening of the features. The SPES spectrum shows a slight non-zero intensity before the ionization threshold attributed to the thermal population of vibrational energy levels of the ground state of the DABCO molecule prior to ionization, which remains from the inefficient cooling of the vibrational degrees of freedom in the expansion. It is assumed that the rovibrational envelope does not drastically change for the corresponding vibrational transition (labeled 0,2 and 4). The Gaussian widths were given boundary conditions based on the widths of the three most clearly resolved transitions. Above $\approx$~7.7~eV, the fit does not carry any physical meaning. The relative intensities of the observed vibronic transitions reflect the Franck–Condon overlap between the ground state of the neutral molecule and the vibrational levels of the cation. At least two vibrational modes contribute to the spectrum, forming two vibrational progressions shown by the grey dashed and yellow dotted Gaussian series.
\begin{table*}[htbp]
\caption{Energies of the vibronic transitions in eV evaluated for three different measurements $\rm A,~\rm B~\text{and}~\rm C$\ (from the best fit to the data presented in Fig.~\ref{fig:SPES}). Double asterisks (**) indicate energies for which the precision is limited due to spectral congestion and are therefore excluded from the rest of the analysis. The first transition (\textbf{0}) is the ionization threshold, evaluated to 7.199~eV~$\pm$~0.006~eV. The numbering of the vibrational transitions is arbitrary and is only used for referencing purpose.}
\label{tab:nu}
\centering
\begin{tabular}{c *{15}{c}}
\toprule
$n~\nu_{25}$ & \textbf{0} & 1 &   & 2 &   & 3 &   & 4 &   & 5** &    & 6** &    & 7** \\
$\nu_{23}~+n~\nu_{25}$ &  & & 0' &   & 1' &   & 2' &   & 3' &   &4'** & &  5'** &&  \\
\midrule
$\rm A$ & 
7.201 & 7.304 & 7.355 & 7.403 & 7.464 & 7.511 & 7.562 &
7.607 & 7.664 & 7.717 & 7.771 & 7.828 & 7.882 & 7.945 \\
$\rm B$ &
7.197 & 7.299 & 7.353 & 7.400 & 7.461 & 7.506 & 7.556 &
7.599 & 7.664 & 7.713 & 7.753 & 7.801 & 7.858 & 7.903 \\
$\rm C$ &
7.199 & 7.300 & 7.357 &  &  &  &  &
 &  &  &  &  & & \\
 $\text{mean}$ &
7.199 & 7.301 & 7.355 & 7.402 & 7.463 & 7.509 & 7.559
 & 7.603 & 7.664 & 7.715 & 7.762 & 7.815 & 7.870 & 7.924 \\
&\footnotesize$\pm$0.006 &\footnotesize $\pm$0.006&\footnotesize $\pm$0.006 &\footnotesize $\pm$0.008 &\footnotesize $\pm$0.008& \footnotesize$\pm$0.009 &\footnotesize $\pm$0.009 &\footnotesize $\pm$0.009 &\footnotesize $\pm$0.008 & \footnotesize$\pm$0.009 &\footnotesize $\pm$0.012 &\footnotesize $\pm$0.016 &\footnotesize $\pm$0.015 &\footnotesize $\pm$0.023 \\ 
\bottomrule
\end{tabular}
\end{table*}
Fig.~\ref{fig:SPES}, lower panel, shows the high-resolution spectrum of the DABCO molecule between 7.1~eV and 7.39~eV. The measured photoelectron spectrum is represented in grey and the fitted spectrum in red. The corresponding ion spectrum is overlaid in dark blue. One can notice that similar information can be retrieved from the ion signal and the photoelectron signal. In the ion spectrum, a change in slope marks the population of a new cationic vibrational state. Three step-like features can be identified. The middle portion of each slope, where the slope remains approximately constant, aligns with an individual transition in the photoelectron spectrum. The dashed vertical lines indicate the energies of the vibronic transitions extracted from the data analysis and used in the fitting procedure. The relative intensity of the transition 1 (at~$\approx~7.35$~eV) indicates that at least two vibrational modes contribute to the Franck–Condon envelope of the measured spectrum. A careful examination of the 2D spectrum Fig.~\ref{fig:SPES}, top panel, also corroborates this assumption. No clear evidence for a transition at $\approx$~7.25~eV is observed. The non-zero signal at this energy can safely be attributed to the excitation of low-frequency modes in the neutral ground state that are easily thermally populated \cite{quesada_hot-band_1986,poisson_unusual_2010}, commonly referred to as hot bands.   
\newline
Table~\ref{tab:nu} lists the energies of the vibronic transitions from the ground state of the neutral molecule to the vibrationally excited states of the cation identified in this work. Even-numbered transitions are associated with the first vibronic progression, identified as the yellow dotted Gaussian series in Fig.~\ref{fig:SPES}, while odd-numbered transitions are associated with the second vibronic progression, represented by the grey dashed Gaussians in Fig.~\ref{fig:SPES}. Transitions marked with a double asterisk (**) correspond to energies obtained from the best fit to the data, but are no longer used to interpret the results. Based on these data, we provide a newly determined value of the DABCO adiabatic ionization threshold of 7.199~eV~$\pm$~0.006~eV.
\newline
Franck-Condon progressions (corresponding to the even-numbered and odd-numbered transitions) are used to evaluate the vibrational frequency associated with each identified mode. In the anharmonic case, the quantum mechanical energy levels are given by $E_{\nu} = (\nu + \tfrac{1}{2}) \, \omega_e - (\nu + \tfrac{1}{2})^2 \, \omega_e \, \chi_e $ where \(\omega_e\), and \(\omega_e\chi_e\) are the harmonic frequency and the anharmonicity constant. We use this expression to determine the vibrational frequencies. 

The two vibrational progressions are fitted separately as a function of an arbitrary integer representing the vibrational quantum number. Only the first five (resp. four) vibronic transitions of the first (resp. second) progression are fitted. The resulting vibrational frequency is $\omega_1=833~\text{cm}^{-1}~\pm~19~\text{cm}^{-1}$ with anharmonic constant $3 ~\text{cm}^{-1}\pm~4~\text{cm}^{-1}$ for the first progression. 
The second progression appears shifted from the origin by $\delta=1257~\text{cm}^{-1}~\pm67~\text{cm}^{-1}$, has a vibrational frequency of  $\omega_2=861~\text{cm}^{-1}~\pm~42~\text{cm}^{-1}$ with anharmonic constant $8~\text{cm}^{-1}\pm~11~\text{cm}^{-1}$. These frequencies correspond to the excitation of the same mode, which frequency is given by averaging $\omega_1$ and $\omega_2$. 

In the literature, the vibrational frequencies and optimization of the DABCO molecular structure are often calculated within the harmonic approximation \cite{kovalenko_experimental_2012,nizovtsev_structural_2017}, and our data are consistent with a negligeable anharmonicity. The large uncertainty results from the strong correlation between the fitted parameters \(\omega_e\) and \(\chi_e\), as well as the small number of vibrational levels included in the fit. % thus the anharmonic term is supposed to be close to zero.
The ground state potential energy surface of the DABCO molecule has been found to be either single or double-welled, depending on the level of theory used \cite{nizovtsev_structural_2017}. This double-well is associated with low-frequency twisting modes calculated for the neutral ground state \cite{quesada_hot-band_1986,poisson_unusual_2010}. The energy of the separation level of the two-welled potential is linked to a positive anharmonicity \cite{quesada_hot-band_1986, nizovtsev_structural_2017}. Therefore some degree of anharmonicity is expected.  
\newline
The DABCO molecule has 54 vibrational degrees of freedom, which reduces to 36 normal modes considering the molecular symmetry. In the $\mathrm{D_{3h}}$ point group, the irreducible representation is $\Gamma_{\text{vib}}=6a_{1}'+3a_2'+4a_1''+5a_2''+9e'+9e''$ , where the modes $a_1'$, $e'$ and $e''$ are Raman active, the $a_2''$ and $e'$ are IR actives, and the $a_2'$ and $a_1''$ are optically inactive. The geometry and electronic structure of the radical cation are mostly similar, and it also has 36 normal modes \cite{balakrishnan_radical_2000}. 11 of them have been observed in the gas-phase in a two-color multiphoton ionization experiment \cite{fujii_two-color_1983} and 4 modes have been observed for DABCO in aqueous solution using resonance Raman spectroscopy \cite{balakrishnan_radical_2000}. All of them are computed using Hartree-Fock and DFT methods \cite{balakrishnan_radical_2000}. To the best of our knowledge, only two studies have reported investigations of the vibrational energies of the DABCO cation, namely those of \citet{balakrishnan_radical_2000} and \citet{fujii_two-color_1983}. Thus, we discuss our results in the context of their findings. In particular, the assignment of the cation vibrational modes follows that of \citet{balakrishnan_radical_2000}. 
Upon ionization, the equilibrium geometry of the system changes from that of the neutral molecule to that of the cation. This change in geometry is a multidimensional shift, that sensibly has components along several cation normal modes. This results in more than one cation mode being Franck-Condon active. 

We observe two Franck-Condon progressions with vibrational frequencies $\omega_1$ and $\omega_2$ that are associated with the population of a single mode at $847~\text{ cm}^{-1}$ with an uncertainty of $~\pm~27~\text{cm}^{-1}$. This mode agrees well with the observed frequency at $854~\text{ cm}^{-1}$ for the cation ($\nu_{25}$ with $e'$ symmetry)\cite{balakrishnan_radical_2000}. 

The second progression is shifted by one quantum of vibrational energy of another mode. This means that the second progression corresponds to a combination of two cation modes, likely normal modes. The shifted origin $\delta=1257~\text{cm}^{-1}~\pm67~\text{cm}^{-1}$ is largely consistent with an excitation of the $\nu_{23}$ mode ($e'$ symmetry), measured at $1245~\text{cm}^{-1}$ \cite{fujii_two-color_1983} and calculated at $1235~\text{cm}^{-1}$ \cite{balakrishnan_radical_2000}. 
\noindent A comparison may be made with the corresponding harmonic frequency of the neutral molecule available in \citet{kovalenko_experimental_2012} and \citet{balakrishnan_radical_2000}. Vibrational modes of the neutral molecule with $e'$ symmetry are observed at $890~\text{cm}^{-1}$ and at $1320~\text{cm}^{-1}$in the gas phase by IR spectroscopy \cite{kovalenko_experimental_2012}. Both IR-active modes ($e'$) have a weak oscillator strength. The first one is assigned to an in-plane twisting motion of the methylene groups ($\text{CH}_2$) and antisymmetric stretching of the $\text{C-C}$ bound \cite{kovalenko_experimental_2012}. The second one is associated to an in-plane twisting motion of the methylene groups ($\text{CH}_2$) and antisymmetric stretching of the $\text{N-C}_3$ groups \cite{kovalenko_experimental_2012}. According to the literature, there would be no immediate reason to invoke a possible symmetry change between neutral and the cation. In our data, the Franck–Condon transitions associated with ionization into the cationic potential energy well are relatively strong. The maximum of the vibrational progression envelope is shifted towards 7.5~eV, indicating that the geometric change upon ionization to the cation ground state is moderate. Thus it is reasonable to expect that the vibrational motions in the cation are close to that of the neutral. However, quantitative calculation of the Franck–Condon factors would be required for an unambiguous assignment and description of the excited vibrational modes.

\subsection{\label{sec:Vib-level2} Photoelectron Anisotropy}
Upon one-photo ionization with a linearly polarized light, the photoelectron angular distribution of gas-phase, isotropically distributed molecules is given by 
\begin{equation}
    I(\Theta) \propto  \textbf{P}_0(\text{cos}\Theta) + \beta\textbf{P}_2(\text{cos}\Theta)
\end{equation}
where $\Theta$ is the angle between the ejection direction of the photoelectron and the light polarization axis, $\textbf{P}_2(\text{cos}\Theta)$ is a Legendre polynomial of order 2 and $\beta$ is the anisotropy parameter, constrained to take values between $-1$ and $2$. 
% so that the distribution remains non-negative. 
An anisotropy parameter of $2$ corresponds to a pure dipole-like emission, where the distribution is entirely aligned with the polarization axis. The case where $\beta$ is close to $1$ corresponds to an intermediate anisotropy, where the photoelectron emission is forward-backward along the polarization axis, but not exclusively. An anisotropy parameter close to zero indicates an isotropic photoelectron angular distribution. Within the Born–Oppenheimer and Franck–Condon approximations, the anisotropy parameter is determined by the electronic transition dipole moment between the initial and final electronic states. Under these approximations, one does not expect the anisotropy parameter to vary upon excitation of different vibrational levels within the same electronic state. However several studies have reported that $\beta$ parameters can depend on the vibrational excitation.
\newline
$\text{H}_2\text{O}$ has been investigated by resonant Auger decay, where the variation of $\beta$ with vibrational excitation within a given electronic state was attributed to the interference between direct and resonant ionization pathways. The lifetime of the resonantly excited state was also shown to influence the scattering of the outgoing electron \cite{hjelte_angular_2005}. Similar conclusions were drawn for $\text{CO}$, where direct-resonant pathway interference and vibrational excitation of the intermediate resonant state were proposed to explain the observed dependency of the $\beta$ parameter with the vibrational excitation \cite{kukk_auger_1999}. A more recent resonant Auger photoemission study on $\text{O}_2$ showed that the dependence of the anisotropy parameter on the final vibrational state arises from direct–resonant interference and non-adiabatic valence–Rydberg mixing in the core-excited state \cite{lindblad_vibrational_2012}. 

\begin{figure}[htbp]
    \centering
    \includegraphics[width=0.4\textwidth]{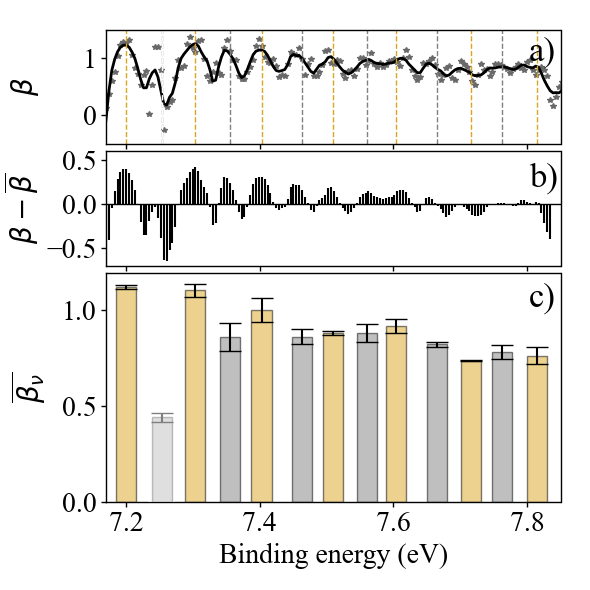}
    \caption{\label{fig:Anisotopy} a): Photoelectron anisotropy parameter. The grey stars are the data, the black thick line is the filtered data with a Savitsky-Golay filter with a span of 11 and a polynomial of order 3. The dashed lines indicate the position of the two Franck-Condon progressions listed in Table.~\ref{tab:nu}. b): Deviation from the mean of the anisotropy parameter over the whole range between the threshold and $\approx$~7.85~eV, calculated at $\bar{\beta}=~$0.83. c): Mean photoelectron anisotropy parameter, obtained by averaging the data over a $\pm~15~\text{meV}$ interval around the transitions. Error bars represent the standard deviation between measurements \textbf{A} and \textbf{B} (see Table.~\ref{tab:nu}). The signal between the first two transitions of the first progression (shaded in the figure) is excluded from the analysis. The \textit{x}-axis is corrected by the Stark shift of respectively $6.21~\rm meV$ and $7.55~\rm meV$ for the two measurements used in this figure.}
\end{figure}

In photoelectron spectroscopy studies of $\text{N}_2\text{O}$, \cite{ferrett_vibrationally_1989} $\text{N}_2$ \cite{dehmer_shape-resonance-enhanced_1979} and $\text{CO}$ \cite{siggel_resonance_1992, semenov_interplay_2004} the dependence of the anisotropy parameter with the vibrational excitation was attributed to shape resonances. Shape resonances are quasi-bound states that arise from the temporary trapping of photoelectrons by a centrifugal barrier in the effective potential, which distorts the continuum wavefunction. Shape resonances strongly depend on the nuclear geometry. The study of branching ratios near shape resonances evidences strong coupling between electronic and vibrational motion, leading to pronounced non–Franck–Condon behavior. In such cases, the dipole transition amplitude depends on the nuclear geometry, and therefore indirectly on vibrational excitation. Autoionizing state have also been reported to influence the angular distribution \cite{siggel_resonance_1992}, likely through interference with the direct ionization pathway. 

To the best of our knowledge, only two studies on larger molecular systems report vibrational anisotropy. Acetylene has been investigated by resonant Auger emission spectroscopy. The vibrational anisotropy was attributed to direct-resonant pathway interference and to interference between outgoing electron wavefunctions scattered through different regions of the double-well core-excited potential accessible via bending motion (Renner–Teller effect). In this case, the transition to the resonant state depends on the nuclear coordinates. Thus, this constitutes a breakdown of the Franck-Condon approximation. Finally, methyloxirane was studied by valence photoelectron spectroscopy using circularly polarized light \cite{garcia_vibrationally_2013}. The vibrational dependence of the anisotropy parameter was attributed to a breakdown of the Franck-Condon approximation. 

The ground state highest occupied orbital in DABCO is a $\sigma$-type nonbonding orbital centered on the nitrogen's pair. A one photon ionization with a linearly polarized light is expected to yield a positive anisotropy parameter. Starting from the raw matrix, an inversion is performed to obtain two distinct distributions: the energy distribution corresponding to $\rm P_0$ and that associated with the angular distribution $\rm P_2$, for the full range of photon energies. These distributions are then integrated over a 300 meV window in photoelectron energy, along the cationic states, which allows constructing the $\beta$ parameter using the SPES method, by taking the ratio $\rm P_2 / \rm P_0$. The anisotropy parameter as a function of the photon energy is shown between 7.17~eV and 7.85~eV in Fig.~\ref{fig:Anisotopy}. Beyond this range, the low signal intensity leads to significant divergence of the $\beta$ parameter. The data (grey stars) is overlaid by a Savitsky-Golay filtered data, using a span of 11 and a polynomial of order 3 (black line), used for visualization purpose. The position of the vibrational transitions as listed in Table. \ref{tab:nu} is indicated with the yellow and grey dotted lines for the first and second progression respectively. The photoelectron anisotropy parameter exhibits clear oscillatory behavior as a function of the excited vibrational mode. Panel b) shows the deviation from the mean anisotropy parameter $\bar{\beta}=~0.83$ over the considered range. A clear decrease of the anisotropy parameter is observed with increasing vibrational excitation of the first mode (yellow dashed lines). Panel c) shows the mean anisotropy parameter, evaluated by averaging the data over a $\pm~15~\text{meV}$ interval around each transition. Integration intervals are defined based on the peak FWHM obtained from the fitting procedure presented above. This value was chosen to ensure the determination of the anisotropy parameter for individual vibrational transitions with sufficient statistics. For the first vibrational progression (yellow dashed lines and bars), the mean anisotropy parameter $\bar{\beta_2}$ decreases from approximately 1.1 at threshold to about 0.8 with higher vibrational transitions. For the second vibrational progression (grey dashed lines and bars), the anisotropy parameter remains constant with a value of $0.83~\pm~0.04$. The averaged anisotropy parameters $\bar{\beta_\nu}$ for each transition are given in Table. \ref{tab:beta}. The average of the ratio $\rm P_2 / \rm P_0$ over a given interval is not equivalent to the ratio of the averages of $\rm P_2$ and $\rm P_0$ calculated separately over the same interval. Fluctuations in $\rm P_0$, especially when it takes small values or contains noise, amplifies certain contributions. In contrast, taking the ratio of the averages implicitly weights the contributions, leading to a more stable estimate. The values in the table were obtained by averaging the $\beta$ obtained from the two methods The corresponding standard deviation was calculated by taking into account the standard deviations from both methods, effectively reflecting the larger of the two uncertainties.

\begin{table}[b]
\centering
\caption{Photoelectron anisotropy parameters obtained by averaging the data within $\pm~15~\text{meV}$ around each transition.}
\label{tab:beta}
\begin{tabular}{cccccccc}
\toprule
 & 0 & 1 & 2 & 3 & 4 & 5 & 6 \\
\midrule
\textbf{$\beta_\text{I}$} & 1.12 & 1.1  & 1  & 0.88  & 0.91  & 0.77  & 0.81 \\
\textbf{$\sigma_\text{I}$} & 0.01 & 0.04 & 0.06 & 0.01 & 0.04 & 0.03 & 0.01 \\
\midrule
 &  & 0' & 1' & 2' & 3' & 4' & \\
\midrule
\textbf{$\beta_{\text{II}}$} &  & 0.86 & 0.84 & 0.88 &  0.79 & 0.79 & \\
\textbf{$\sigma_\text{II}$} &  & 0.07 & 0.05 & 0.05 & 0.02 & 0.04 & \\
\bottomrule
\end{tabular}
\end{table}

The vibrational anisotropy observed for the ground cationic state of the DABCO molecule could be attributed to the contribution of the numerous high-lying Rydberg states near and above the ionization threshold. In the DABCO molecule, four well resolved Rydberg series were found in the energy region between the adiabatic and vertical ionization threshold found at 7.31~eV \cite{fujii_two-color_1984}. These states are autoionizing states, where the autoionization mechanism was found to be vibrational ionization \cite{fujii_two-color_1984}. 

\begin{figure}[htbp]
    \centering
    \includegraphics[width=0.42\textwidth]{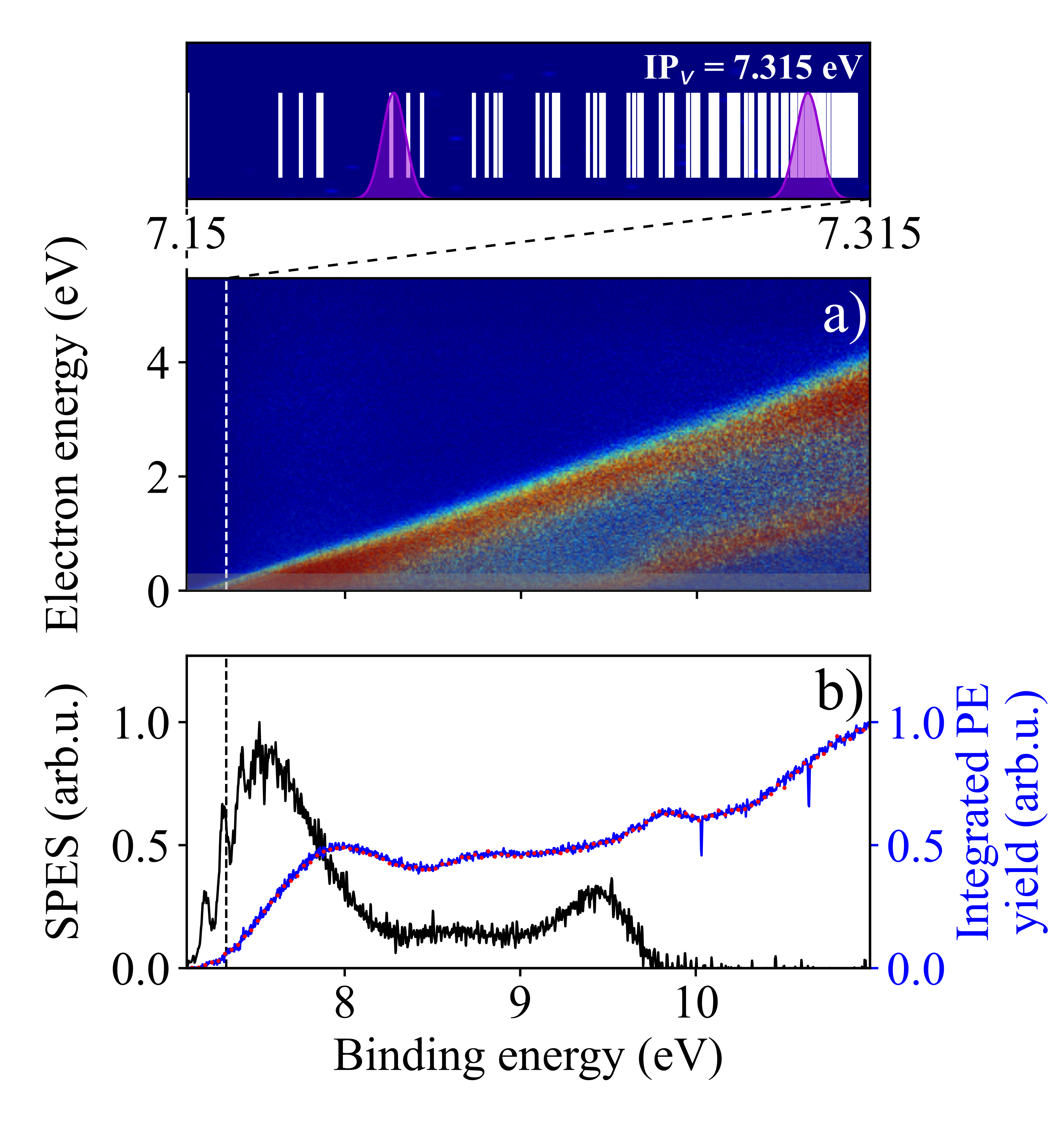}
    \caption{\label{fig:Rydberg}~a) Photoelectron spectrum of the DABCO molecule between 7.1~eV and 11~eV. Inset: expanded view of the 88 Rydberg states line positions from. \cite{fujii_two-color_1984} The synchrotron energy resolution is schematically represented by the violet Gaussian-shaped profile (FWHM of 3~meV, corresponding to the highest energy resolution in our data). b) SPES of the DABCO obtained by integration over the first 300~meV (black line) shown with the integrated photoelectron yield over all the photoelectron energies (dark blue line) and corresponding ion yield (red dashed line). The \textit{x}-axis is corrected by the Stark shift of $10.76~\rm meV$.}
    \end{figure}

Fig.~\ref{fig:Rydberg}~a) shows the overall photoelectron spectrum recorded. The integrated photoelectron yield is plotted Fig.~\ref{fig:Rydberg}~b) (dark blue line). The red-dashed line is the corresponding ion yield. The light grey box in the photoelectron spectrum marks the photoelectron energy range used to obtain the SPES spectrum, displayed in grey Fig.~\ref{fig:Rydberg}~b). The line positions of the 88 Ryberg states listed in \cite{fujii_two-color_1984} are shown in the expanded view in Fig.~\ref{fig:Rydberg}.~a). The energy resolution of the synchrotron radiation is represented by the violet Gaussian-shaped profile, with a FWHM of 3~meV, corresponding to the highest energy resolution achieved in our data. The broadening at higher photon energy is barely perceptible. Given the high density of Rydberg states is high in this region of the molecular potential and our experimental energy resolution, it is clear that multiple states within this energy range were populated, depending on their cross-section.
Interference between the direct ionization pathway and the vibrational autoionization pathway is likely to affect the photoelectron angular distribution, i.e., the amplitude and phase of the continuum wavefunction. Moreover, Rydberg states lying above the adiabatic ionization potential could be seen as quasi-bound states embedded in the continuum. Their coupling to the ionization continuum modifies the continuum wavefunction and lead to resonance features in the scattering of the outgoing electron. The contribution from Rydberg states suggested in this interpretation might also indicate a non-Franck-Condon picture for ionization from the neutral ground state to the cation. In \citet{briant_reaction_2022}, the cationic states of the $\text{Ar}_2^+$ are populated indirectly following relaxation from Rydberg states, while the cationic states themselves do not have oscillator strength as a result of a strong change in molecular geometry upon ionization.
\newline
At higher photon energies, additional Rydberg series converging toward excited states of the ionic core may contribute to the signal. The first excited electronic state of the cation is reached at $9~\text{eV}\pm0.4~\text{eV}$, which corresponds to the threshold of the second band in the photoelectron spectrum. An increase in the photoelectron signal associated with a given ionic state can reflect an enhanced excitation or ionization probability (e.g. a higher photoionization cross-section).
However, in Fig.~\ref{fig:Rydberg}, the increase of the signal associated with the ionic ground state (first band) is accompanied by a broadening towards lower kinetic energy electrons. This behavior is also observed in the integrated photoelectron yield, which shows broad intensity increases with maxima around 8.0~eV, 8.8~eV, 9.8~eV, and above 10.5~eV. The broad structures indicate the population of short-lived autoionizing states strongly coupled to the ion core. They are consistent with the presence of Rydberg states undergoing vibrational autoionization, belonging to series converging toward the first and second excited states of the ionic core. This interpretation is further supported by the weak signal of slow photoelectrons above the first excited cationic state. Similar previous studies on coronene reported a Rydberg series converging to the fourth excited cationic state \cite{brechignac_photoionization_2014}. In this work, the relatively sharp autonization features in the photoelectron spectrum (0.1~eV) indicates the presence of long-lived Rydberg states, and correspondingly a weak coupling with the ion core.
\newline
Therefore, not only the Rydberg series converging towards the ionization threshold (adiabatic or vertical), but also those converging towards at least the first excited state of the ionic core should be considered when interpreting the observed vibrational anisotropy.

\section{Conclusions}
We presented a photoelectron spectroscopy study of the DABCO molecule performed at the DESIRS beamline of the SOLEIL synchrotron. The high-resolution, high-flux, and tunable photon source allowed us to vibrationally resolve the ground state of the cation. Two vibrational progressions were observed. The first was assigned to the $\nu_{25}$ mode of $e'$ symmetry, while the second was assigned to a combination of the $\nu_{23}$ and $\nu_{25}$ normal modes. The assignment was based on structural considerations of the corresponding motions in the neutral molecule, as well as on previously observed and calculated frequencies for the DABCO cation. In addition, we have performed a complete analysis of the associated photoelectron angular distribution and observed an unexpected dependence of the anisotropy parameter with the vibrational excitation in the cationic ground state. We attribute this effect to possible interference between the direct ionization pathway and autoionizing high-lying Rydberg states, which exist up to 0.1~eV above the ionization threshold. These Rydberg states can also be considered quasi-bound and may influence the scattering of the outgoing electron. Rydberg series converging towards higher excited states of the ionic core might as well have contributed to the observed anisotropy. 
Such a behavior would not a priori be anticipated for a highly symmetric molecule ionized with linearly polarized light. The role of Rydberg states on the photoelectron angular distribution in larger molecular system could be systematically studied.

\section{Author contributions}
L. P. has administered the project incl. writing the project proposal and conceptualization. L. P. developed the software (LABVIEW program) to extract and process all relevant experimental parameters. A. S. is the main investigator incl. data analysis, conceptualization and manuscript writing. B. G., G A. G., L. N. took part in the experiment, discussion and contributed to reviewing the manuscript. E. G. and L.B. took part in the experiment and contributed to reviewing the manuscript. D. C. took part in the experiment. C. S. contributed to reviewing the manuscript. All authors approved the final version of the manuscript.

\section{Conflicts of interest}
There are no conflicts to declare.

\section{Data availability}
All data are available from the corresponding author upon reasonable request.

\vspace{2em}
\section{Acknowledgements}
This research was funded in part by Agence Nationale de la Recherche (ANR) under the project "ANR-23-CE29-0012-03". We acknowledge SOLEIL for provision of synchrotron radiation facilities and we would like to thank J.-F. Gil for assistance in using beamline DESIRS (project \#20231511). 

\vspace{8em}
\twocolumngrid
\bibliography{MainBib}
\bibliographystyle{BibStyle} 
\end{document}